\def \s{~\rm{s}}
\def \km{~\rm{km}}

\def \K{~\rm{K}}

\def \AU{~\rm{AU}}

\def \yr{~\rm{yr}}

\documentclass[]{emulateapj}

\shorttitle{On the Upper PNLF}
\shortauthors{Soker}
\slugcomment{Draft version of \today}
\begin{document}

\title{ACCRETING WHITE DWARFS AMONG THE PLANETARY NEBULAE MOST LUMINOUS
IN [O~III]~5007 EMISSION}
\author{Noam Soker \altaffilmark{}}
\affil{Department of Physics, Technion$-$Israel Institute of Technology,
Haifa 32000 Israel;
soker@physics.technion.ac.il}
%\altaffiltext{1} { }

\begin{abstract}
I propose that some of the most luminous planetary nebulae (PNs) are actually
proto-PNs, where a companion white dwarf (WD) accretes mass at a relatively
high rate from the post-asymptotic giant branch star that blew the nebula.
The WD sustains a continuous nuclear burning and ionizes the nebula.
The WD is luminous enough to make the dense nebula luminous in
the [O~III]~$\lambda5007~$\AA~ line,
In young stellar populations these WD accreting systems
account for a small fraction of [O~III]-luminous PNs,
but in old stellar populations these binaries might account for most,
or even all, of the [O~III]-luminous PNs.
This might explain the puzzling constant cutoff (maximum) [O~III]~$\lambda 5007$
luminosity of the planetary nebula luminosity function
across different galaxy types.

\end{abstract}
%{\bf Key words:}

% ==========================================================
\section{INTRODUCTION}
\label{sec:intro}
% ==========================================================
The planetary nebula (PN) luminosity function (PNLF) depends on the age
of the parent population (Dopita et al. 1992; Mendez et al. 1993;
Marigo et al. 2004), and to lesser degree on the metallicity
(Ciardullo \& Jacoby 1992; Marigo et al. 2004),
hence it varies between different types of galaxies (e.g.,
Ciardullo et al. 2004).
However, the [O~III]~$\lambda5007~$\AA~ (hereafter [O~III])
most luminous end of the PNLF, and in particular the cutoff (maximum) [O~III]
luminosity, seems to be similar in all large PN populations,
with very small dependance on galaxy type (Ciardullo et al. 2005).
This allows a very successful use of the PNLF as a standard candle,
from the galactic bulge (Pottasch 1990), through the LMC and SMC
(Jacoby et al. 1990), M31 (Ciardullo  et al. 1989)
and to galaxies at larger distances, spirals (e.g.  Feldmeier et al. 1997),
and ellipticals (Jacoby et al. 1996).

Some basic properties of the PNLF are well understood (e.g., Dopita et a. 1992;
Mendez et al. 1993; Marigo et al. 2004; review by Ciardullo 2006).
However, a major puzzle is the cutoff (maximum) luminosity in old stellar populations,
such as in elliptical galaxies, which is the same as in young populations
(Ciardullo et al. 2005; Ciardullo 2006).
This cutoff luminosity of the PNLF is $L$[O~III]~$\simeq 600 L_\odot$,
and it leads to a chain of constraints
(Ciardullo et al. 2005; Ciardullo 2006):
Such [O~III] luminosity requires the ionizing central star to have a bolometric
luminosity of $L_\ast \ga 6000 L_\odot$, which in turn requires the
central star to have a mass of $M_\ast > 0.6 M_\odot$, and the progenitor
to have a main sequence mass of $M> 2 M_\odot$.
Such a progenitor mass is not expected in old stellar populations.
{{{ Single star evolution alone cannot account for this finding (Marigo et al. 2004). }}}

Ciardullo et al. (2005) propose that the most [O~III] luminous PNs in old populations
are descendant of blue-straggler type stars; namely, two lower mass
stars, $\sim 1 M_\odot$, merged on the main sequence to form a star of
mass $\sim 2 M_\odot$.
Such a population of luminous PNs is quite possible.
I  propose another plausible type of [O~III]-luminous PNs,
composed of a post-AGB star and a close accreting white dwarf (WD)
companion.

% ==========================================================
\section{OBSERVATIONAL HINTS}
\label{sec:obs}
% ==========================================================
Several types of observations hint at the possibility that some of the
[O~III]-luminous PNs are actually proto-PN systems ionized by an
accreting WD companion.

{\it (1) Morphology:}
{{{ Throughout this paper it is assumed that the bipolar morphology of
PNs is acquired via binary interaction (Soker 1998).
I stress that the stellar binary model for shaping bipolar PNs is a conjecture,
and that it is not accepted by all researchers in the field. }}}
In the local galactic PNLF according to Mendez et al. (1993) four of the five
most luminous PNs are bipolar, NGC 7027 (Latter et al. 2000), NGC 7026
(Corradi \& Schwarz 1995); NGC 6572 (Miranda et al. 1999) and
NGC 6537 (Corradi \& Schwarz 1995). Only 7009, fourth most luminous,
is elliptical but with two jets along the symmetry axis (Balick 1987).
We note that NGC 7027, the most luminous by a large margin in the Mendez's sample,
is a near by compact PN, only recently turned into a bipolar PN (Kastner 2005).
At larger distances, several kpc or more, its bipolar structure
would have been unresolved, and it would have been classified as
a round or an elliptical PN.
Therefore, the fact that many [O~III]-luminous PNs in the Magellanic clouds
are classified round or elliptical (Magrini et al 2004; Stanghellini 2005)
is not surprising, and cannot be used as a contra observation to the claim
above.
Some spherical [O~III]-luminous PNs can be formed from massive single stars.
However, I expect many of the [O~III]-luminous PNs that are classified
round and elliptical to have inner bipolar structure, as in NGC 7027.
Blue stragglers are not expected to form bipolar PNs (Soker 1998)
unless a close tertiary star exists, but this is very rare.
{{{ It should be noted that in the galaxy and the Magellanic clouds
there are young stellar populations, and most of the [O~III]-luminous PNs
can be explained by single star evolution (Marigo et al. 2004).
The point here is that I expect a large fraction of the most [O~III]-luminous
PNs in these galaxies to have a binary companion, with some of the companions being
accreting WDs. }}}

{\it (2) Young PNs:}
The central star of NGC 7027 left the AGB $\sim 700$ years ago
(Latter et al. 2000).
The bipolar PN NGC 7026 is at a similar evolutionary stage
(Cuesta et al. 1996) with a dynamic age of $\sim 800$ years
(Solf \& Weinberger 1984).
The age of the luminous bipolar PN NGC 6572 is similar to
that of these two luminous PNs (Miranda et al. 1999).
These young PNs suggest that many [O~III]-luminous PNs are young.
Interestingly, NGC 7027 has a relatively high extinction
(Robberto et al. 1993; Kastner et al. 2002), showing that a PN does
not need to clear all its surrounding dust before becoming a luminous PN.
In the Milky Way galaxy, unlike in elliptical galaxies with old stellar populations,
there are massive progenitors that can explain the most [O~III]-luminous PNs.
However, even in these a fast evolution is required for the PN to
spend a long enough time period in an ionized bounded phase.
In the ionized bounded  phase all ionizing radiation is absorbed by the nebula;
hence, the formation of the [O~III] line is more efficient.
A binary companion can disturb the AGB envelope, causing high mass loss
rate and fast post-AGB transition to the PN stage.

{\it (3) Luminous supersoft X-ray sources (SSS):} Many of the SSS
are WDs accreting mass and sustaining continuous nuclear burning.
A large fraction of these WDs have a luminosity of $L_{\rm WD}
\sim 10^4 L_\odot$, and a temperature of $ T_{\rm WD} \sim 1-2
\times 10^5 \K$ (e.g., Di Dtefano \& Kong 2004), appropriate for
the formation of [O~III]-luminous PNs (Dopita et al. 1992). We
note that in the WD-accreting model proposed in this paper the WDs
accrete at higher rates than most SSSs. However, because the soft
X-rays of SSSs are readily absorbed by material shed by the
post-AGB primaries, SSSs in close binary systems with post-AGB
primaries are likely to go undetected in X-rays.
{{{ It should be noted that in the present model {\it close} means
orbital separation of $\sim 1 \AU$, and not $\sim {\rm few} \times R_\odot$.
Hence, the binary systems that will formed from such systems will not merge
to form type Ia SNs. }}}

{\it (4) Central close binary systems in PNs:}
Some binary post-AGB stars demonstrate that binary systems at an orbital
separation of $\sim 1 \AU$, with main sequence or WD companions, can survive without
entering a common envelope phase (Van Winckel 2004).
The companion leads to the formation of a bipolar nebula, as in
the Red Rectangle (Cohen et al. 2004).
The presence of a disk around the Red Rectangle and similar systems suggests that
this phase can last hundreds of years and longer.

{\it (5) Symbiotic systems evolving to PNs:}
There are several similarities between symbiotic nebulae and bipolar PNs
(Corradi \& Schwarz 1995; Kwok 2003), causing some misclassifications.
However, many of the symbiotic systems are evolving to become PNs.
In a PN, by definition, the star that ionizes the nebula is the star that
blew the nebula.
In symbiotic nebulae the secondary ionizes the nebula as they accrete mass
from the primary mass-losing giant star.
{{{ Schwarz \& Corradi (1992) suggested that the bipolar nebula BI Crucis
is an object evolving to become a PN, where there is a central accreting WD.
The type of objects discussed here are similar to their model of BI Crucis.  }}}
In M2-9 the primary star is most probably a post-AGB star (Lim \& Kwok 2000),
while the nebula is ionized by an accreting secondary WD
(Goodrich 1991; Schwarz et~al.\ 1997).
Close examination suggests that M2-9 is a symbiotic nova type system, and
not a PN (Solf 2000).
However, had it been located at a farther distance, such as the galactic bulge,
it would have most probably been classified as a PN.
This classification would be even more probable was M2-9 located in external galaxies.
As the mass losing star begins to ionize the nebula, this system most likely will go
through a smooth transition to a bona-fide PN.

Based on the observations listed above, I propose that some of the [O~III]-luminous
PNs are binary systems where one of the stars, the primary, is a post-AGB star,
while the secondary star is a WD accreting from the primary and sustaining
a continuous nuclear burning.
Such a system is actually a symbiotic system, because the ionizing source is
the companion to the mass-losing star.
However, at this stage of evolution and at the large distances of the galactic
bulge and external galaxies, the dense ionized nebula has all the
characteristics of PNs (Kwok 2003).
As suggested by the observations of near by proto-PN systems, this phase can
last for hundreds of years.
These systems are similar to M2-9 but the orbital separation is
$\sim 1 \AU$, like that in the Red Rectangle,  rather than $\sim 27 \AU$
as in M2-9 (Livio \& Soker 2001).
The much smaller orbital separation than that in M2-9 is required
for a much higher accretion rate by the WD (sec. 3.2).
The high mass accretion rate is required to sustain a continuous nuclear
burning, causing the system to become an SSS.
However, because of the relatively high accretion rate,
the WD swells and becomes somewhat cooler, and because the post-AGB star loses
mass at a higher rate than the mass losing star in most SSSs, the SSS is
highly obscured.
In external galaxies these types of systems most likely will not
be classified as SSSs.

% ==========================================================
\section{MAIN PROPERTIES OF THE PROPOSED MODEL}
\label{sec:evolution}
% ==========================================================

\subsection{Relatively high mass accretion rates}
To sustain a continuous nuclear burning the WD accretion rate should be higher than
some critical value which increases with WD mass; this critical mass
accretion rate is
$\dot M_{{\rm WD}c} \simeq 8 \times 10^{-8} M_{\odot} \yr^{-1}$,
for a WD mass of $M_{\rm WD}=0.6 M_\odot$,
$\dot M_{{\rm WD}c} \simeq 1.6 \times 10^{-7} M_{\odot} \yr^{-1}$
for $M_{\rm WD}=0.8 M_\odot$, and
$\dot M_{{\rm WD}c} \simeq 2.5 \times 10^{-7} M_{\odot} \yr^{-1}$, for
$M_{\rm WD}= 1 M_\odot$ (Hachisu et al. 1999).
At an accretion rate of $\dot M_{{\rm WD}} \sim (2-4) \dot M_{{\rm WD}c}$
the accreting WD swells, blows a wind, and its temperature drops from
$T_{\rm WD}>3 \times 10^5 \K$ to lower temperatures (Hachisu et al. 1999).
For example, a temperature of $T_{\rm WD} = 2 \times 10^5 \K$
(this is the temperature of the central star of the PN NGC 7027; Latter et al. 2000)
is reached at accretion rates of
$\dot M_{{\rm WD}} \simeq 1.6 \times 10^{-7} M_{\odot} \yr^{-1}$
for $M_{\rm WD}=0.6 M_\odot$,
$\dot M_{{\rm WD}} \simeq 6 \times 10^{-7} M_{\odot} \yr^{-1}$ for $M_{\rm WD}=0.8 M_\odot$,
and $\dot M_{{\rm WD}} \simeq 10^{-6} M_{\odot} \yr^{-1}$ for
for $M_{\rm WD}=1 M_\odot$.
As temperature drops toward $T_{\rm WD} \sim 1.2 \times 10^5 \K$ the formation
of the [O~III] line becomes more efficient (Dopita et al. 1992).

\subsection{Close binary systems}
For the WD to accrete at a high enough rate, the orbital separation cannot be
larger than a few AU.
Consider the post-AGB phase and scale the central star
post-AGB stellar wind parameters following the properties of M2-9.
The primary mass loss rate drops already to
$\dot M_{\rm AGB} \sim  {\rm few} \times 10^{-6} M_\odot \yr^{-1}$,
but the post-AGB envelope is still large with a wind speed of
$v_w \sim 20 \km \s^{-1}$.
With the orbital speed of $\sim 30 \km \s^{-1}$,
the WD-wind relative speed is $v_{\rm rel} \sim 35 \km \s^{-1}$.
The WD accretion rate from the wind is
\begin{eqnarray}
\dot M_{\rm WD} \simeq 10^{-6}
\left(\frac {\dot{M}_{\rm AGB}}{3 \times 10^{-6} M_\odot \yr^{-1}}\right)
\left(\frac {M_{\rm WD}}{M_\odot}\right)^2
\nonumber \\
\times \left(\frac{a}{\AU} \right)^{-2}
\left(\frac{v_w}{20 \km \s^{-1}}\right)^{-1}
\left(\frac{v_{\rm rel}}{50 \km \s^{-1}}\right)^{-3} M_\odot \yr^{-1}.
\label{dotm}
\end{eqnarray}
This implies that the WD accretes a large fraction of the
post-AGB wind.
The luminosity is $L_{\rm WD} > 10^{4} L_\odot$, with
part of the accreted mass being blown by the WD, and the WD swells to
a larger radius (Hachisu et al. 1999).
For a given accreting WD sustaining a continuous nuclear burning the
luminosity is constant at
$L_{\rm WD}=5000 L_\odot$ for $M_{\rm WD}=0.6 M_\odot$,
$L_{\rm WD}=10^4 L_\odot$ for $M_{\rm WD}=0.7 M_\odot$, and
$L_{\rm WD}= 3 \times 10^4 L_\odot$ for $M_{\rm WD}=1 M_\odot$
(van den Heuvel et al. 1992).
% More massive WDs are much luminous, but we note the following.
% First, most WDs have masses of $\sim 0.6 M_\odot$, and only a small fraction
% has masses higher than $0.65 M_\odot$.
% Second, the more massive WDs are hotter, hence less efficient at
% producing the [O~III] line despite their much higher luminosity.

Considering the critical mass accretion rate given in \S3.1,
equation (\ref{dotm}) shows that the orbital separation must be $a \la 1 \AU$.
Such binary systems must experience strong binary interaction, namely, strong
tidal interaction, and/or Roche lobe overflow (RLOF), and/or a
common envelope interaction.

\subsection{Bipolar morphology}
For the [O~III] line not to be obscured by dust, a collimated fast wind (CFW)
or jets blown by the accreting WD should clear a path with lower optical depth,
as in the model for M2-9 proposed by Livio \& Soker  (2001).
As evident from the case of NGC 7027 (\S2) it is enough
that extinction is low in only part of the nebula.
Indeed, as discussed in connection to M2-9, the accreted mass has a high
enough specific angular momentum and an accretion disk is formed around the WD.
Such a disk is assumed to blow jets.
In the case of a common envelope interaction, the formation of a bipolar nebula
requires the WD to blow jets before or after the common envelope phase (Soker 2004).
An accretion disk around the WD will be formed via RLOF.
An accretion disk is likely to be formed, and blow jets, even when accreting
from the AGB or post-AGB wind.
The condition for the formation of an accretion disk around the WD can be expressed in the
form $R_d\ga R_{\rm WD}$, where $R_d$ is the outer radius of the disk that
would form given the wind's specific angular momentum.
The disk radius is given approximately by (e.g.\ Livio \& Warner 1984;
see Livio \& Soker 2001 for the full expression)
\begin{equation}
R_d \sim R_\odot \left( \frac{a}{1 \AU} \right)^{-3}
\left({{M_{\rm WD}}\over{{\rm M}_{\odot}}}\right)^3
\left({{v_{\rm rel}}\over{50~{\rm km~s}^{-1}}}\right)^{-8}~{\rm cm}.
\end{equation}
An $M_{\rm WD}=0.6 M_\odot$ WD swells to a radius of $R_{\rm WD} \la 1 R_\odot$ as long as
the accretion rate is $\dot M_{\rm WD} \la 2 \times 10^{-6} M_\odot \yr^{-1}$.
For higher mass WDs the mass accretion rate should be even higher.
We conclude that for the conditions assumed here, where the mass loss rate from the
primary already dropped from its maximum value on the upper AGB, an accretion
disk will be formed, leading to jets (or a CFW) and the formation of a bipolar PN.

\subsection{Mass loss rates}
Still, for the CFW from the WD to clear a path for the ionizing radiation
and ionizing a large volume of the nebula, the primary post-AGB mass loss rate
cannot be too high; as seen below, the constraint is
$\dot M_{AGB} \la 10^{-5} M_\odot \yr^{-1}$.
At the same time the mass loss rate should be
$\dot M_{AGB} \ga 10^{-6} M_\odot \yr^{-1}$, depending on the orbital separation
and WD mass, to sustain continuous nuclear burning on the accreting WD.

Following the analysis of Livio \& Soker (2001) of M2-9,
I compare the nebular recombination rate with ionizing photon emission rate.
The total recombination rate from a minimum distance $r_{\rm min}$ to
large distances in the nebula $r_l \gg r_{\rm min}$ is given by (assuming a spherical flow)
\begin{eqnarray}
\dot R = 4 \pi \int_{r_{\rm min}}^{r_l} \alpha n_i n_e r^2 dr \simeq
5 \times 10^{47}
\nonumber \\
\times
\left( {{\dot M_{\rm AGB}} \over {10^{-6}~{\rm M}_\odot \yr^{-1} }}
\right)^2
\left( {{v_w} \over {20 \km \s^{-1} }} \right)^{-2}
\left( {{r_{\rm min}} \over {1 \AU }} \right)^{-1}
\s^{-1},
\label{rec}
\end{eqnarray}
where $\alpha$ is the recombination coefficient, $n_e$ and $n_i$ are
the electron and ion  number densities (respectively). Here we assumed
a fully ionized nebula of solar composition.
The number of ionizing photons emitted by the WD per unit time, assuming a
temperature of $T_{\rm WD} \simeq 2\times10^5 \K$, is
\begin{equation}
\dot S \simeq 4 \times 10^{47}
\left( {{L_{\rm WD}}\over{10^4 L_\odot}} \right) \s^{-1}.
\label{dots}
\end{equation}

The two rates obtained above, for $\dot R$ and $\dot S$, are within an
order of magnitude from each other, implying that the WD can ionize the nebula
to moderate distances in cases of a spherical geometry.
Jets (or a CFW) which are expected to be blown in the WD-accreting model,
can clear a paths through the dense slow wind to a distance of
$\sim 3-10$ times the orbital separation.
Namely, ${r_{\rm min}} \sim 3-10  \AU $, hence the recombination rate is much lower
than that given by equation (\ref{rec}).
In this bipolar geometry the nebula can be ionized to larger distances than
in the spherical geometry, even for mass loss rate as high as
$\dot M_{\rm AGB} \simeq 10^{-5} M_\odot \yr^{-1}$.
The extended nebula regions are less obscure by dust, enabling the escape
of the [O~III] line.

\subsection{Rapid post-AGB evolution}
As mentioned in \S3.2 the constraint of high accretion rate from a star
after its AGB peak mass loss rate implies a close binary system.
Such strongly interacting WD companions cause large disturbances in the post-AGB
envelope of the primary star, leading to an enhanced
mass loss rate and a fast transition from the AGB to the PN phase.
{\it However, while the WD is accreting at a high rate, it is the WD secondary that
ionizes the nebula, not the primary. }
Therefore, by definition this is a symbiotic like system and not a PN. From the
observational point of view, it might be classified as a young PN, as M2-9 is.
Because of the fast evolution and small orbital separation, the primary will
start ionizing the nebula shortly after the WD ceased to ionize it, or even
before the WD ceased to ionize the nebula.
When the primary star ends with a low mass core, such as when the progenitor is of
low initial mass, the most luminous phase occur when the accreting WD ionizes the
nebula, and not the PN phase itself.

\subsection{Statistics.}

Ciardullo et al. (2005) suggested that the progenitor of most [O~III]-luminous
PNs in old stellar populations are blue stragglers, with a mass about twice the
main-sequence turnoff mass.
They estimate that these PNs spend $\tau_b \simeq 500 \yr$ in the
luminous stage.
With this value of $\tau_b$ they estimate that $\sim 5 \%$ of stars should
become [O~III]-luminous PNs.
They then suggest that the fraction of blue stragglers in old-population
galaxies is well suited to the proposition that all [O~III]-luminous PNs
in these galaxies result from blue stragglers.
In young stellar population galaxies, such as the Milky Way, most
[O~III]-luminous PNs comes from massive stellar progenitors.
One of the problems I see in Ciardullo et al. (2005) estimation is that most
blue stragglers have masses below twice the main sequence turnoff mass
(e.g., Lee et al. 2003; Piotto et al. 2004; De Marco et al. 2005),
with less than $\sim 10 \%$ having a mass more than 1.7 times the turnoff mass.
{{{ The final core mass of the PN descendants of these blue stragglers, though, will
be higher than that of main sequence star of the same mass. }}}
It is not clear that there are enough blue stragglers to
explain the [O~III]-luminous PNs in old-population galaxies.

Soker \& Rappaport (2000) estimated that in the Milky Way galaxy
$\sim 1\%$ of PNs have been formed in a binary system where
the secondary was a strongly interacting WD, as required
for the WD-accreting model proposed in the present paper.
This percentage could be somewhat larger, because Soker \& Rappaport
did not evolve systems which had strong interaction on the
first giant branch (RGB); some of these might eventually also form PNs
progenitors with strongly interacting WD secondary star.
Also, in old stellar populations the fraction of WD companions is expected
to increase, as the massive components of binary systems have already gone
through the PN phase.
Therefore, it seems that PNs progenitors with strongly interacting WD companions
can account for $\sim 1-3 \%$ of the PN population.
The fraction of the systems discussed here among all observed PNs can
be even higher if the vast majority (Soker \& Subag 2005), or even all
(De Marco \& Moe 2006), PNs are formed by binary stars.
 
In addition, the accreting WD companion can stay luminous for longer
than the $\tau_b = 500 \yr$ assumed by Ciardullo et al. (2005).
{{{ As an example consider again M2-9. The accreting WD companion is
currently still ionizing the nebula (Goodrich 1991; Schwarz et~al.\ 1997).
The dynamical age of the outer nebula is $\ga 1200$~yr
(Schwarz et al. 1997; Solf 2000), suggesting the the binary
interaction, including high mass transfer rate,
can last for $>1000$~yrs.
Schwarz et al. (1989) suggest that He 2-104 is both a proto-PN and
a symbiotic system. They estimate the nebular dynamical age to be
800~yrs. These studies of M2-9 and He 21-104 show that a system
can be a symbiotic type, with a high mass loss rate from the
primary simultaneously with ionization by an accreting WD
companion, for $> 1000$~yrs.
This time scale is compatible with that expected in the model.
A post-AGB star shrinks to $\sim 50 \%$ its maximum AGB size
when its envelope mass is $\sim 3 \times 10^{-3} M_\odot$, while
the rapid shrinking occurs when the envelope mass is
$\sim 10^{-3} M_\odot$ (Soker 1992).
This implies that the post-AGB star loses
$\sim {\rm few} \times 10^{-3} M_\odot$ during the relevant phase. From the
discussion in \S3.4 the mass loss rate is required to be
$\sim {\rm few} \times 10^{-6} M_\odot \yr^{-1}$.
According to the proposed model, therefore, the relevant phase lasts
$\sim 1000~$yrs. }}}   From this discussion it seems quite plausible
that the strongly interacting WD systems proposed in this paper comprise
most, or even all, of the  [O~III]-luminous PNs in galaxies with an
old stellar population.
This might account for the finding that the cutoff luminosity
of the PNLF in such galaxies is the same as that of galaxies with
a much younger stellar population.

% ==========================================================
\section{SUMMARY}
\label{sec:summary}
% ==========================================================

Motivated by observations described in \S2,
I proposed that some of the [O~III]-luminous PNs are actually proto-PNs,
where a WD companion accretes mass at a relatively high rate from the post-AGB
star, sustains continuous nuclear burning, and ionizes the dense nebula.
Such systems are actually symbiotic type nebulae, similar to
supersoft X-ray sources in many respects.
However, because of the dense nebula and large distances to the
external galaxies, these are classified as PNs.
In young stellar populations these strongly-interacting WD binaries
account for a small fraction of [O~III]-luminous PNs,
{{{  and single star models can account for the PNLF (e.g., Marigo et al. 2004). }}}
However, in old-stellar populations, such as in many elliptical galaxies,
these binary PNs might account for most, or even all, of the
[O~III]-luminous PNs (\S3.6).
This could be the solution to the puzzling constant cutoff (maximum)
[O~III]~$\lambda 5007$ luminosity of the PNLF across different galaxy types.
{{{  The constant cutoff [O~III]~$\lambda 5007$ luminosity is accounted for
by the behavior of the accreting WDs:
($i$) Like for single central stars of PNs (Marigo et al.  2004)
most WDs are concentrated around mass $M_{WD} \sim 0.6 M_\odot$,
with very small number having $M_{\rm WD}>0.7 M_\odot$.
($ii$) For a given WD, the luminosity does not increase above a maximum value
(van den Heuvel et al. 1992), which is
$L_{\rm WD}=10^4 L_\odot$ for $M_{\rm WD}=0.7 M_\odot$.
($iii$) For higher luminosities the WD accretion rate should be higher,
requiring a higher mass loss rate by the post-AGB star.
This in turn implies that the ionized region is small and obscured.
These three properties ``conspires'', I propose, to give a constant cutoff
[O~III]~$\lambda 5007$ luminosity, somewhat similar to the explanation
of single stars models (Marigo et al. 2004).
I note that the luminosity of these systems declines monotonically as the mass
loss rate from the post-AGB star drops, hence accretion rate decreases.
Later the post-AGB central star heats-up enough to ionize the nebula, and
luminosity can rise again.
These systems, therefore, fill the entire bright part of the PNLF, and
merge with the distribution of single stars. }}}

This model predicts that in old stellar populations the [O~III]-luminous PNs
are bipolar, despite the low mass of their progenitors.
However, because PNs are [O~III] luminous when they are young, hence small,
it is very challenging to spatially resolve such PNs.
As some fraction of [O~III]-luminous PNs in young stellar populations
are also expected to be proto-PNs with accreting WDs, it will be interesting
to search for such WDs inside [O~III]-luminous PNs in our galaxy.

\acknowledgments This research was supported in part by the Asher
Fund for Space Research at the Technion.

\end{document}